# Single-frequency lasers' linewidth elegantly characterized with Sigmoid functions of observation time

XIAOSONG MA,[1] AND X. STEVE YAO[1,2,*]

[1] *Photonics Information Innovation Center and Hebei Provincial Center for Optical Sensing, College of Physics Science and Technology, Hebei University, Baoding 071002, China.*
[2] *NuVison Photonics, Inc, Las Vegas, NV 89109, USA.*
*\* syao@ieee.org*

**Abstract:** Linewidth is the most important parameter for characterizing the coherence properties of a single-frequency laser, but unfortunately only the natural linewidth representing the contributions of the spontaneous emission or quantum noise can be described with an analytical expression known as the Schawlow–Townes–Henry formula. To the best of authors' knowledge, no analytical expression is formulated after 63 years since laser's invention for characterizing the effective linewidth of a single-frequency laser including the linewidth broadening caused by the flicker noises, which strongly depends on the measurement duration and is much larger than the natural linewidth. By carefully measuring the instantaneous frequency fluctuations of multiple commercial single-frequency lasers using a self-built optical frequency analyzer with ultra-high resolution and speed to obtain their linewidths with our time domain statistical analysis method, we discover and validate that the laser linewidths can be expressed as one or more Sigmoid functions of observation time. Not only the simple Sigmoid linewidth expression provides clear linewidth information of the laser, but also better understanding of the physical origins affecting the laser linewidths, which will benefit a large number of applications ranging from coherent distributed sensing to gravitational wave detection and therefore is worthy to be widely adopted to fully and elegantly characterize the linewidths of single-frequency lasers.

## 1. Introduction

Single-frequency lasers are commonly used in a wide range of applications, including coherent optical communication [1, 2], high-resolution spectroscopy [3, 4], coherent Lidar [5], fiber optic sensing [6-8], optical atomic clocks [9, 10], and gravitational wave detection [11]. The linewidth of a laser generally refers to the full width at half maximum (FWHM) of the laser's power spectral density and is one of the most important parameters for characterizing the performance of single-frequency lasers. The effective linewidth of a single-frequency laser generally includes the natural linewidth caused by the spontaneous emission noise (white noise) and the technical linewidth resulting from the flicker noises (pink noise or 1/f noise) due to laser cavity fluctuations of various types [12, 13]. The natural linewidth is the lower limit of the laser linewidth and can be expressed by the Schawlow–Townes–Henry (STH) formula [14, 15], while the technical linewidth is considered as the broadening from this lower limit, which generally increases with the measurement or observation time [16] and can be hundreds times larger than the natural linewidth. In coherent communication systems, the natural linewidth determines the coherent property of the laser because the flicker noise is negligible over a symbol interval [17] on the order of ns or less. On the other hand, in most sensing applications, it is this effective linewidth that determines the actual performance parameters of the systems containing the single-frequency lasers, such as the measurement range, resolution, noise, detection speed, and sensitivity [18-24], because the signal measurement time is on the order of *μ*s or more. Therefore, the ability to quickly and accurately determine the effective linewidth of single-frequency lasers at different observation times is of paramount importance for sensor system designs. Unfortunately, for over 65 years since the Schawlow–Townes limit [14] for the natural linewidth was reported, no analytical expressions have been identified or formulized to fully characterize the effective linewidth of single-frequency lasers against the observation time, although cumbersome numerical methods are available to calculate the effective linewidth from the measured frequency noise data [25, 26].

Many schemes have been utilized to measure the linewidths of single-frequency lasers [27], including the heterodyne method [28-31] by beating the laser under test (LUT) with a reference laser, the self-heterodyne method by beating the LUT's output with a Brillouin laser pumped by the LUT itself [32], the delayed self-homodyne method by beating the LUT output with its delayed replica [33], and the delayed self-heterodyne method by first frequency-shifting the LUT's output with an acousto-optic modulator (AOM) and then beating it with the non-frequency shifted LUT output going through a long delay [17, 34, 35]. The beat signal in each scheme is then analyzed by an electrical spectrum analyzer (ESP) to obtain the power spectral density (PSD) of the laser light field $S_E(f)$, for further linewidth analysis. In order to relax the need of requiring ultra-long optical fiber for providing sufficiently large delay in the self-heterodyne schemes, fiber re-circulating loop methods [36, 37] and PSD envelope fitting methods [38, 39] have also been developed. However, all of these schemes are mostly used for obtaining the natural linewidths of the LUT's, not the effective linewidth containing the contributions of the flicker noise which strongly depend on the measurement time. In fact, the flicker noise contributions to the linewidth measurement are purposely excluded by the Voigt fitting scheme in practice in order to obtain more accurate natural linewidth results [17, 40, 41].

In order to obtain the effective linewidth of a single-frequency laser containing the contributions of the flicker noise, schemes of directly obtaining laser's frequency noise $\Delta\nu(t)$ was proposed and implemented [25, 42, 43]. One method of directly obtaining $\Delta\nu(t)$ is to use an unbalanced interferometer based on a 3×3 coupler [43, 44]. As an approximation, after the power spectral density $S_{\Delta\nu}(f)$ of $\Delta\nu(t)$ is calculated, a so called β-separation line is plotted on the $S_{\Delta\nu}(f)$ vs. frequency curve to find the intersect points with the PSD curve [25]. The laser's effective linewidth can be obtained by calculating the area of $S_{\Delta\nu}(f)$ above the intersect points [42]. In principle, the corresponding observation time may be determined by selecting the lower bound frequency of $S_{\Delta\nu}(f)$ data in the integral for calculating the area, although the previous works [25, 42, 43] did not report or mention such a capability.

It is important to mention that all the linewidth measurement schemes mentioned above analyze data in the spectral domain, manifested by the needs of processing the data to get either the PSD of the lasers' light field [$S_E(f)$] or the PSD of the lasers' frequency fluctuation [$S_{\Delta\nu}(f)$], which is complex and time consuming because the resolution of the spectrum is limited by the sampling time and multiple averaging is required to obtain the spectral accuracy.

In this paper, we take a different approach to analyze the data for obtaining the laser linewidth. We first measure the laser frequency fluctuations with time, $\Delta\nu(t)$, using one of the sine-cosine frequency detection schemes [45-48] with sufficiently high frequency resolution and sufficiently high speed for a sufficiently long period. Instead of the spectral domain analysis, we analyze laser frequency fluctuations $\Delta\nu(t)$ statistically in the time domain, without the need of obtaining the spectral domain PSD [$S_{\Delta\nu}(f)$]. The laser linewidths corresponding to different measurement durations or observation times $T_{ci}$ can be obtained by calculating the probability distribution function (PDF) of $\Delta\nu(t)$ after $\Delta\nu(t)$ being filtered by high-pass filters of different cutoff frequencies $f_{ci}$ ($T_{ci} = 1/f_{ci}, i = 1 \ldots N$), and then obtaining the width of each PDF by curve fitting to a Voigt function, which is then taken as the linewidth of the laser. With this approach, we can quickly obtained the laser linewidth as a function of the observation time $T_c$.

We have measured multiple commercially available single-frequency lasers using this approach and have discovered that the linewidths of these lasers can be fitted as the sum of one or more Sigmoid functions of the observation time, with specific parameters to represent the minimum linewidth (the natural linewidth), the maximum linewidth, and the slope rate of the linewidth change in between. To the best of authors' knowledge, this is the first time that the laser linewidth vs. observation time can be fully described by an analytical expression.

It is interesting to note that the S shaped Sigmoid function is commonly used in deep learning architectures as a nonlinear activation function to guarantee the output is always between two states, 0 and 1 [49, 50]. Here the Sigmoid function guarantees that the laser linewidth is between two fixed values, determined by the observation time. The lower limit is determined by the white noise dominant over short observation times, while the upper limit is determined by the flicker noise dominant over long observation times. The case of needing more than one Sigmoid functions of observation time to represent the linewidth indicates that an additional process is involved, such as a frequency stabilization or relative intensity noise (RIN) reduction loop in the laser system, to affect the laser's linewidth. Therefore, the Sigmoid presentation of the linewidth can actually unveil the noise generation/reduction processes involved inside the laser system.

Finally, we have verified that the laser natural linewidths obtained with our time domain linewidth analysis (TDLA) method follow the well known STH formula as the laser power is varied. More importantly, the Sigmoid linewidth expression can also be validated by the linewidth values obtained with the well accepted spectral domain linewidth analysis (SDLA) method, particularly the β-separation line method.

Our work fills a long standing vacancy for an analytical expression to accurately describe the linewidth of a single-frequency laser as a function of observation time, which is important for applications involving interferences or requiring laser coherence properties.

## 2. The laser frequency noise measurement and time domain linewidth analysis

### 2.1 Optical frequency detection system

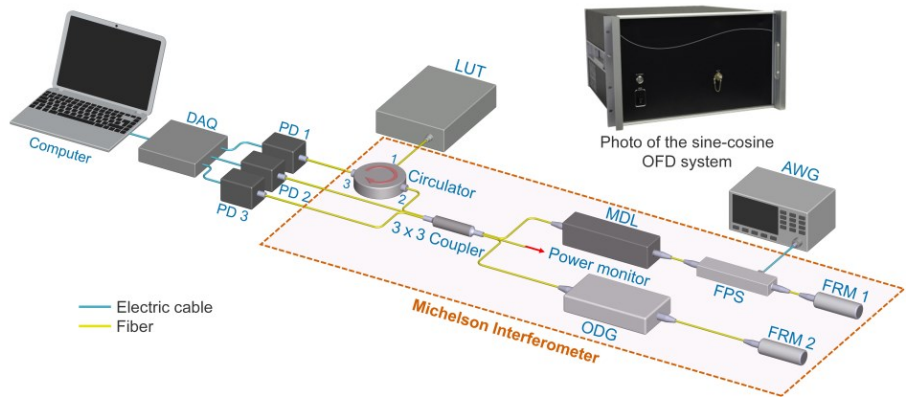

Fig. 1. Schematic of a cosine-sine OFD based on a Michelson interferometer with a 3×3 coupler. PD: photodetector; MDL: motorized variable delay line; ODG: programmable optical delay generator; FPS: Fiber phase shifter; AWG: Arbitrary waveform generators; FRM: Faraday rotator mirror; DAQ: data acquisition card (16 bits resolution with up to 1Gs/s sampling rate).

The self-built sine-cosine optical frequency detection (OFD) system is based on a unbalanced Michelson interferometer constructed with a 3×3 coupler, as shown in Fig. 1, which is one of the sine-cosine OFD schemes described in [47-48] with extremely high frequency resolution and high detection speed. Similar setup was used previously for measuring laser's frequency noises [43]. The LUT is first split into three arms by a 3×3 coupler after passing through a circulator. Two of the three arms are used to form a Michelson interferometer with a 90° Faraday mirror at the end of each arm to get rid of polarization fluctuations [51], and the other beam is used for monitoring the optical power. A motorized variable delay line (MDL) is added to one of the interferometer arms to provide continuous optical path difference (OPD) variations up to 400 mm. A programmable optical delay generator (ODG) [52] is added to the other arm to provide eight different OPDs for the interferometer: 23, 114, 205, 296, 385, 476, 566, and 658 m. The delay in the interferometer determines the frequency measurement resolution and needs to be selected according to the linewidth of the laser being measured. When the linewidth is large, a short OPD is required in the interferometer to avoid exceeding the measurement speed limit due to the instantaneous frequency change rate of the laser. When the linewidth is small, a large OPD is required in the interferometer to obtain a higher system resolution. A fiber phase shifter (FPS) made with a piezo-electric actuator is placed in one of the interferometer arms, which can be used to slightly modulate the OPD for calibrating the OFD system, as will be discussed shortly.

The two beams reflected by the Faraday mirrors interfere in the 3×3 coupler to output three beams, ideally having a phase difference of 120° with one another. The three beams are directed into three photodetectors (PD1, PD2, and PD3), respectively, with beam 1 being first routed by the circulator, to convert them into three photovoltages, $V_1(t)$, $V_2(t)$ and $V_3(t)$. The whole interferometer in Fig. 1 is carefully insulated thermally in an enclosure with high density polytetrafluoroethylene material, which also serves to damp acoustics and mechanical vibrations.

In general, the three photovoltages can be expressed as [53, 54]

$$V_1(t) = C_1 + B_1 \cos[\Delta\theta(t) + \beta_1] \tag{1a}$$
$$V_2(t) = C_2 + B_2 \cos[\Delta\theta(t) + \beta_2] \tag{1b}$$
$$V_3(t) = C_3 + B_3 \cos[\Delta\theta(t) + \beta_3] \tag{1c}$$

where $C_i$ ($i = 1,2,3$) are the offset coefficients resulting from the DC bias of the electronic circuit, $B_i$ are the amplitude coefficients of the measurement system relating to the coupling ratios of the coupler, the responsivity of the photodetectors, and amplification factors of the corresponding circuit, $\beta_i$ ($i = 1,2,3$) are the phases of three light beams due to the coupler, and $\Delta\theta(t)$ is the phase change due to the laser frequency variation $\Delta v$, which relates to the time delay $\tau$ of the OPD by:

$$\Delta\theta(t) = 2\pi\tau\Delta v(t) \tag{2}$$

For a perfect 3×3 coupler and ideally balanced photodetectors and associated amplification circuit, $C_i = C$, $B_i = B$, $\beta_1 = \beta_0$, $\beta_2 = \beta_0 - 120°$, $\beta_3 = \beta_0 + 120°$, $\Delta\theta(t)$ can be expressed as a sine and a cosine function of the three photovoltages, and be further solved unambiguously by taking the ratio of them for a tangent function without phase wrapping issues [47]. It can be shown from Eqs. (1a) – (1c) that even in the non-ideal situations, $\Delta\theta(t)$ can still be solved with a tangent expression:

$$\tan\Delta\theta(t) = \frac{\left(\frac{B_3}{B_1}\cos\beta_3 - \frac{C_3}{C_1}\cos\beta_1\right)*\left[V_2(t) - \frac{C_2}{C_1}V_1(t)\right] - \left(\frac{B_2}{B_1}\cos\beta_2 - \frac{C_2}{C_1}\cos\beta_1\right)*\left[V_3(t) - \frac{C_3}{C_1}V_1(t)\right]}{\left(\frac{C_2}{C_1}\sin\beta_1 - \frac{B_2}{B_1}\sin\beta_2\right)*\left[V_3(t) - \frac{C_3}{C_1}V_1(t)\right] - \left(\frac{C_3}{C_1}\sin\beta_1 - \frac{B_3}{B_1}\sin\beta_3\right)*\left[V_2(t) - \frac{C_2}{C_1}V_1(t)\right]} \tag{3}$$

The coefficients $C_i$ and $B_i$, as well as the phase differences ($\beta_2 - \beta_1$, $\beta_3 - \beta_1$, $\beta_3 - \beta_2$) can be obtained by using a tunable laser to scan the laser frequency with a sufficiently large range, getting the three voltages $V_1(t)$, $V_2(t)$ and $V_3(t)$, plotting the Lissajous figures of $V_1(t)$ vs. $V_2(t)$, $V_2(t)$ vs. $V_3(t)$, and $V_1(t)$ vs. $V_3(t)$, and finally performing elliptical fits of the these three Lissajous figures [55]. In the procedure above, $\beta_1$ can be assumed to be 0 for simplicity because the three phases are relative. The elliptical fitting procedure for obtaining these coefficients can also be performed actively by modulating the OPD with the embedded FPS or the MDL, or passively by waiting for the temperature variation to modulate the OPD with an amplitude of more than the wavelength of the laser while the laser frequency is fixed. In our measurements reported in this paper, the passive temperature modulation method is used for simplicity. The relative change of the LUT's center frequency $\Delta v(t)$ can be obtained as:

$$\Delta v(t) = \frac{1}{2\pi\tau}\Delta\theta(t) \tag{4}$$

In order to measure lasers with extremely narrow linewidths, the measurement resolution of $\Delta v(t)$ must be sufficiently fine. For example, for measuring the laser linewidth of 100 Hz, $\Delta v(t)$ measurement resolution must be much better than 100 Hz. As discussed in [47], for our system with a maximum OPD of 658 m and a data acquisition card (DAQ) with a resolution of 15 effective bit, a frequency resolution of 7 Hz can be achieved. More detailed discussion on the frequency resolution and other limitations of the sine-cosine OFD can be found in the Supplementary Information section of this paper.

*2.2 Time domain statistical linewidth analysis (TDLA)*

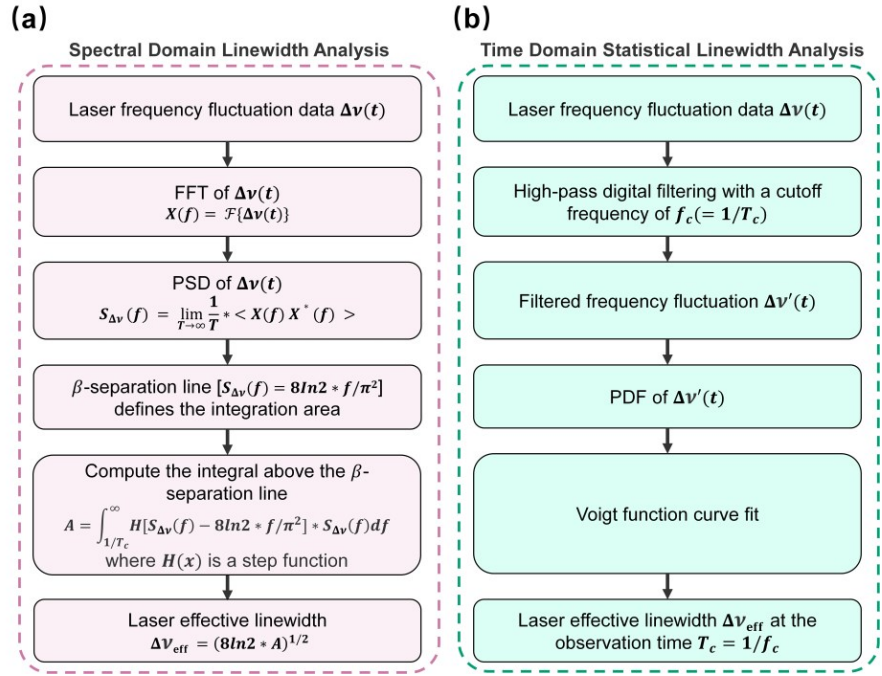

Fig. 2. Comparison of the data processing procedures between the previous SDLA method (a) and our TDLA method (b).

As mentioned in the introduction, previous laser linewidth analyses have most been in the spectral domain. As an example, Fig. 2(a) shows such a SDLA data processing process of using the β-separation line scheme [25, 42, 43] for obtaining the linewidth, in which optical frequency noise data $\Delta v(t)$ of Eq. (4) is first used to compute $S_{\Delta v}(f)$, the PSD of $\Delta v(t)$. Next, the β-separation line is used to divide $S_{\Delta v}(f)$ into two areas, one above the line and the other below the line. Finally, the laser line width can then be determined by calculating the area above the β-separation line.

In contrast, we obtain the laser linewidth using the TDLA approach, as shown in Fig. 2(b). In particular, the laser frequency noise data $\Delta v(t)$ is first processed digitally with a high-pass filter (HPF) having a cutoff frequency $f_c$ to remove slow frequency variations and obtain filtered frequency noise $\Delta v'(t)$, with which the probability density function (PDF) of the frequency noise is computed. Finally, the PDF is curve-fitted to a Voigt function to obtain its -3dB linewidth. By changing the cutoff frequency $f_c$, the laser linewidth at different observation time $T_c$ can be obtained, recognizing that $T_c = 1/f_c$. In comparison with the SDLA method, our TDLA method is more straight forward to implement and faster to compute.

In reality, the measured $\Delta v(t)$ data contains not only the laser's instantaneous frequency fluctuations responsible for the laser line broadening, but also the noises of the measurement system, including the noises from the PD, the amplifiers and the DAQ card, which are not related to the linewidth and should be excluded from the linewidth calculation.

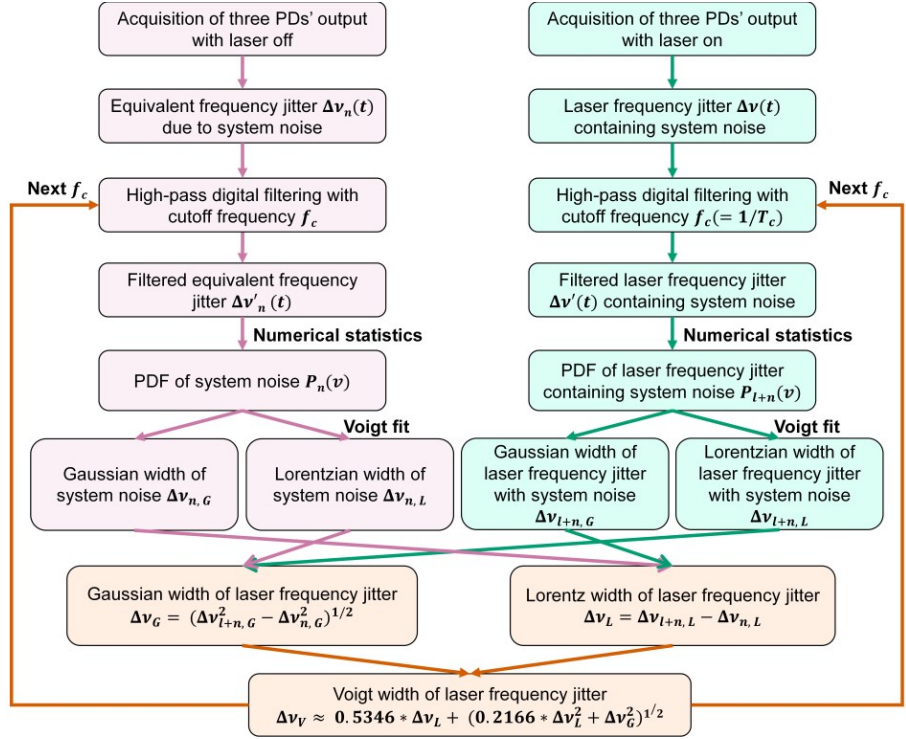

Fig. 3. Flowchart for calculating the effective laser linewidth in the experiment. Collect the data when LUT is off and the data when LUT is on and pass the data through a high-pass filter. Use Voigt line shape fitting to separate the Gaussian linewidth and Lorentzian linewidth of the filtered data, and use the characteristics of line width to eliminate the influence of system noise.

The process of subtracting out the system noise is shown in Fig. 3. One first measures the system noise by taking the data of the three voltages $\delta V_{n1}(t)$, $\delta V_{n2}(t)$ and $\delta V_{n3}(t)$ for a period of time when there is no light input. Next, turn on the LUT and take data of the three voltages as $V'_1(t)$, $V'_2(t)$ and $V'_3(t)$ for the same period of time, which include the system noise contribution and can be expressed as

$$V'_j(t) = V_j(t) + \delta V_{nj}(t), \ j = 1,2,3 \tag{5a}$$

where $V_j(t)$ can be viewed as the detected photo-voltages without the contributions of the system noise. The data set of the three voltages is then used to compute $\Delta\theta(t)$ and hence the laser frequency fluctuation $\Delta\nu(t)$ using Eqs. (3) and (4), which includes the contributions of the system noise.

The system noise contribution to the measured optical frequency fluctuations can be evaluated by first low-pass filtering $\Delta\theta(t)$ obtained above to remove the contributions from the fast laser frequency jitter and system noise such that the photo-voltages in Eq. (1) can be expressed as

$$\widetilde{V}_j(t) = C_j + B_j \cos[\widetilde{\Delta\theta}(t) + \beta_j], \; j = 1,2,3 \tag{5b}$$

where the "~" sign stands for low-pass filtering. The corresponding photo-voltages due to the system noises can be written as

$$V'_{nj}(t) = \widetilde{V}_j(t) + \delta V_{nj}(t) - \overline{\delta V_{nj}(t)}, \; j = 1,2,3 \tag{5c}$$

where the bar stands for time average and $\overline{\delta V_{nj}(t)}$ is a small constant offset. The data set of the three voltages in Eq. (5c) can be used in Eqs. (3) and (4) to calculate the equivalent phase angle $\Delta\theta_n(t)$ and frequency fluctuation $\Delta\nu_n(t)$ due to the system noise alone, respectively. An example of obtaining $\Delta\nu_n(t)$ is included in the Supplementary Information section.

After filtering $\Delta\nu(t)$ and $\Delta\nu_n(t)$ with a 3rd-order Butterworth high-pass filter having a cutoff frequency $f_c$, $\Delta\nu'(t)$ and $\Delta\nu'_n(t)$ are obtained, which can be used to compute the PDFs of the combined laser frequency fluctuation and system noise [denoted as $P_{l+n}(\nu)$], and the system noise alone [$P_n(\nu)$], respectively.

We found in the experiments that each of $P_{l+n}(\nu)$ and $P_n(\nu)$ can be best fitted with a Voigt function, which is the convolution of Gaussian and Lorentzian line shapes [56]. The resulting FWHM width $\Delta\nu_{n,V}$ for $P_n(\nu)$ is called the system noise equivalent linewidth (SNEW). By performing the Voigt fitting and following the procedure in [56], the Gaussian width and Lorentzian width in $P_{l+n}(\nu)$ and $P_n(\nu)$ can be separated. Consequently, the contribution of the system noise to the linewidth measurement can be subtracted. Assuming the laser line shape contains both Gaussian and Lorentzian line shape contributions, the corresponding Gaussian linewidth $\Delta\nu_G$, the Lorentzian linewidth $\Delta\nu_L$, and the final Voigt linewidth $\Delta\nu_V$ of the laser frequency noise PDF at the full width half maximum (FWHM), excluding the contribution of the system noise, can be obtained as [57]

$$\Delta\nu_G = \sqrt{\Delta\nu_{l+n,G}^2 - \Delta\nu_{n,G}^2} \tag{6a}$$

$$\Delta\nu_L = \Delta\nu_{l+n,L} - \Delta\nu_{n,L} \tag{6b}$$

$$\Delta\nu_V \approx 0.5346 * \Delta\nu_L + \sqrt{0.2166\Delta\nu_L^2 + \Delta\nu_G^2} \tag{6c}$$

where $\Delta\nu_{l+n,G}$ and $\Delta\nu_{l+n,L}$ are the Gaussian and Lorentzian linewidths in $P_{l+n}(\nu)$, and $\Delta\nu_{n,G}$ and $\Delta\nu_{n,L}$ are the Gaussian and Lorentzian linewidths in $P_n(\nu)$. This Voigt linewidth is taken as the effective linewidth $\Delta\nu_{\text{eff}}$ of the LUT.

By change the cutoff frequency $f_c$ of the digital high pass filter, the effective linewidths of the LUT with different observation time $T_c$ ($T_c = 1/f_c$) can be obtained using the same procedure described above, as shown in Fig. 3. Due to the limited processing speed of the computer, a fixed number of data points of 10 million (10M) samples is taken by the DAQ card and used for the linewidth computation in our measurements. Therefore, the measurement duration and the sampling speed of the DAQ card must be balanced, which can be controlled by the computer in Fig. 1. Higher sampling speed results in shorter measurement duration, and vice versa. For some lasers, two or more sets of data with different sampling rates must be taken and combined to cover the adequate measurement range.

## 3. Experimental results

### 3.1 3D presentation of the frequency jitter of two commercial lasers

Fig. 4(a) shows the measured frequency jitter $\Delta\nu(t)$ of an external cavity laser at 1550 nm (Pure Photonics PPCL550) as a function of time, which is taken with a sampling rate of 100 Ms/s for a duration of 0.1 s and thus contains 10M data points. The delay length in the OFD (Fig. 1) is chosen to be 658 m, corresponding to a frequency measurement resolution of 7 Hz, which is sufficiently smaller than the 10 kHz linewidth specified by the laser manufacturer.

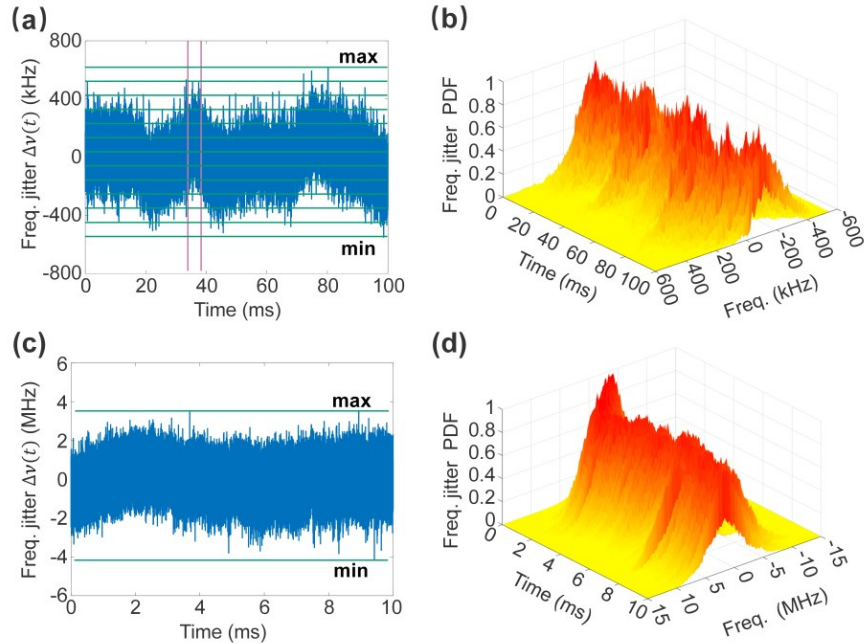

Fig. 4. (a) Measured frequency jitter $\Delta\nu(t)$ of the external cavity laser taken with a sampling rate of 100Ms/s and an OPD of 658 m; (b) the corresponding 3D PDF plot, where color and height represent the amplitude of the PDF. (c) Measured frequency jitter $\Delta\nu(t)$ of the grating-stabilized diode laser taken with a sampling rate of 1Gs/s and an OPD of 23 m; (d) the corresponding 3D PDF plot.

Fig. 4(a) contains all the laser frequency fluctuation information, however, it is difficult to visualize how the line shape and linewidth evolve with time. Fig. 4(b) is the three-dimensional (3D) PDF plot of the external cavity laser's frequency jitter, which is an intuitive representation of its line shape and linewidth. By first dividing the data along the time axis into $M_t$ slices and then along the $\Delta\nu$ axis into $M_{\Delta\nu}$ slices to form $M_t \times M_{\Delta\nu}$ slots, as shown in Fig. 4(a), the PDF of frequency jitter in each slot can be obtained by taking the ratio of the number of data points in each slot over those in that time slice. In obtaining Fig. 4(b), $M_t$ and $M_{\Delta\nu}$ are taken to be 200 such that each time slice has a duration of 0.5 ms. Finally, the 3D PDF is obtained by stitching the PDFs of all 200 time slices together.

Similarly, Fig. 4(c) is the $\Delta\nu(t)$ of a grating-stabilized diode laser at 1550 nm (built-in laser in the APEX AP2051A spectrometer) with an output power of 5 dBm, which is obtained with a sampling rate of 1 Gs/s for a duration of 0.01 s and also contains 10M data points, while Fig. 4(d) is the corresponding 3D PDF plot obtained with the same procedure as that of Fig. 4(b). Note that the OPD here is reduced to 23 m to void the frequency variation rate $\gamma$ exceeding the maximum detection limit $\gamma_{max}$ of the OFD system [47]. In particular, $\gamma$ obtained by taking the derivative of the data in Fig. 3(c) has a maximum value of 400 THz/s, which exceeded the system limit $\gamma_{max}$ of 228 THz/s for the case of OPD=658 m, as described in the Supplementary Information section. In addition, an OPD of 23 m corresponds to a frequency resolution of 200 Hz [47], still much smaller than the 3 MHz typical laser linewidth specified by the laser manufacturer for valid linewidth measurements.

It can be seen from Figs. 4(b) and 4(d) that in addition to the fast frequency fluctuation, each laser's mean frequency also drifts with time. Such drifts will contribute to the linewidth broadening over relative long observation durations.

It should be noted that the linewidth property of a laser may vary as it warms up from cold start, as we observed in the experiments. In order to obtain consistent measurement results, the laser should be warmed up for more than 30 minutes before measurements.

### 3.2 Effective linewidth of lasers as a single Sigmoid function of observation time

Using the TDLA procedure discussed in Fig. 3, the PDFs of the frequency variations of the external cavity laser and the grating-stabilized diode laser are obtained and shown with the blue lines in Figs. 5(a) and 5(b), respectively, in which the cutoff frequency of the high-pass filter is set at 1MHz, corresponding to an observation duration of 1 $\mu$s. In obtaining the PDFs in Fig. 5, the range of laser frequency fluctuation $\Delta v(t)$ is first divided into 5000 equal slices, similar to that in Fig. 4(a), and the probability of the laser frequency falling in each slice is calculated by taking the ratio of the number of data points in that slice over the total number of data points (10M). A Voigt function is used to fit the PDF of the frequency noise of each laser for separating it into a Gaussian and a Lorentzian line shape with a Gaussian and a Lorentzian linewidth.

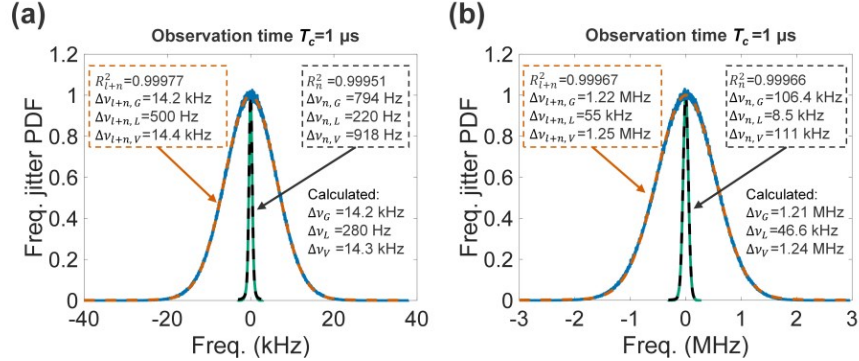

Fig. 5. PDF's of two different lasers with an observation duration $T_c$ of $1\mu s$ ($f_c = 1MHz$). (a) The external cavity laser measured with a sampling rate of 100Ms/s and an OPD of 658 m. (b) The grating-stabilized diode laser measured with a sampling rate of 1Gs/s and an OPD of 23 m. The blue line represents the PDF of the laser frequency jitter containing system noise, while the dashed red line represents its Voigt curve fit. The green line represents the PDF of the system noise, while the dashed black line represents its Voigt curve fit.

In general, the frequency jitter data of Fig. 4(a) contains system noise, which can be subtracted by applying the procedure in Fig. 3. In particular, by turning the laser off, the PDF of the system noise can also be obtained, as shown by the green line inside the line shape defined by the combined laser frequency and system noises in Figs. 5(a) and 5(b), which can also be fitted to a Voigt function shown by the black dashed line and be separated into a Gaussian and a Lorentzian line shape with a Gaussian and a Lorentzian linewidth. Finally, they can be separately subtracted from the corresponding Gaussian and Lorentzian linewidths of the PDF of the combined laser frequency and system noises using Eqs. (6a) to (6c). It can be noticed that in Figs. 5(a) and (b), the goodness of each Voigt fit is better than 0.999, indicating that the line shape of the laser frequency noise can be well represented by a Voigt function.

As shown in Fig. 5(a), with an observation time of 1 μs, the Voigt width and the extracted Gaussian and Lorentzian widths of the external cavity laser's frequency fluctuation are 14.4 kHz, 14.2 kHz, and 500 Hz, respectively, which include contributions from the system noise. For the system noise PDF, the Voigt width and the extracted Gaussian and Lorentzian widths are 918, 794, and 220 Hz, respectively. After subtracting the contributions of system noise, we obtain a Voigt width of 14.3 kHz for the PDF of the external cavity laser's frequency fluctuations over an observation time of 1 μs, which is considered as the effective linewidth of the laser at 1 μs.

Similarly, the PDF of the grating-stabilized diode laser's frequency fluctuation with an observation time of 1 μs is shown in Fig. 5(b). The Voigt, Gaussian, and Lorentzian widths are 1.25, 1.22, and 0.055 MHz, respectively, which include the contributions from the system noise. The corresponding Voigt, Gaussian, and Lorentzian widths of the system noise PDF are 111, 106.4, and 8.5 kHz, respectively. Consequently, after subtracting the contributions of the system noise using Eq. (6), we obtain the Voigt width of 1.24 MHz, which is taken as the effective linewidth of the grating-stabilized diode lasers over an observation time of 1 μs.

Note that in Figs. 5(a) and 5(b), the influence of the system noise is quite small and can be effectively ignored. However, in other measurements with lasers of narrow linewidths, the influence of the system noise is comparatively much larger, which necessitates the need for the subtraction of the system noise contribution. More detailed discussions on this issue can be found in the Supplementary Information section.

Also note that the equivalent linewidth $\Delta\nu_{n,V}$ of the system noise in Fig. 5(b) appears to be much larger than that in Fig. 5(a) because the equivalent frequency fluctuation $\Delta\nu_n(t)$ is inversely proportional to the OPD and also depends on the amplitudes and phase of the interferometer signals when the LUT is on, as described by Eqs. (5b) and (5c). However, such a difference will not affect the subtraction of $\Delta\nu_{n,V}$ from $\Delta\nu_{l+n,V}$ following Fig. 3 when computing each laser's linewidth, because $\Delta\nu_{l+n,V}$ and $\Delta\nu_{n,V}$ for each laser measurement are obtained with the same OPD and the interference signals.

By changing the cutoff frequency $f_c$ of the high-pass filter, the effective linewidth $\Delta\nu_{\text{eff}}$ of the external cavity laser at different observation times $T_c = 1/f_c$ can be obtained following the procedure in Fig. 3, with the results shown in Fig. 6. It is interesting to notice from the figure that the change of $\Delta\nu_{\text{eff}}$ with $T_c$ is "S" shaped, having a "lowland" region with linewidths close to a minimum value $\Delta\nu_{\min}$, a "plateau" region with linewidths close to a maximum value $\Delta\nu_{\max}$ and a rapid linewidth ramping region between $\Delta\nu_{\min}$ and $\Delta\nu_{\max}$, which can be fitted to a Sigmoid function in the form of a logistic function with the excellent goodness of fit better than 0.997:

$$\Delta\nu_{\text{eff}}(T_c) = \Delta\nu_{\max} + \frac{\Delta\nu_{\min} - \Delta\nu_{\max}}{1 + (T_c/T_0)^p} \quad (7a)$$

where $T_0$ is the observation time corresponding to the average linewidth $(\Delta\nu_{\max} + \Delta\nu_{\min})/2$, and $p$ is a parameter for determining the linewidth ramping slope with $T_c$. Note that $\Delta\nu_{\text{eff}}$ can also be written as a Boltzmann function, a more familiar form of Sigmoid function:

$$\Delta\nu_{\text{eff}}(t_c) = \Delta\nu_{\max} + \frac{\Delta\nu_{\min} - \Delta\nu_{\max}}{1 + \exp[(t_c - t_0)/\Delta t]} \quad (7b)$$

where $t_c = \ln(T_c)$, $t_0 = \ln(T_0)$, and $\Delta t = 1/p$. For simplicity, we choose to use the logistic function Eq. (7a) to represent the effective linewidth $\Delta\nu_{\text{eff}}$ vs. observation time $T_c$.

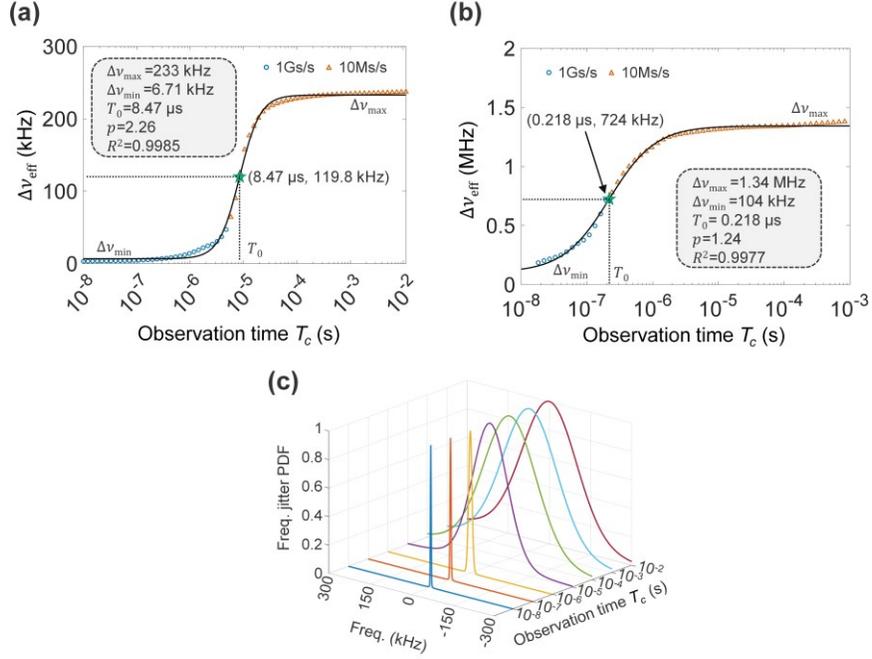

Fig. 6. The effective linewidths of lasers as single Sigmoid function of observation time. (a) External cavity laser with logistic function fit (OPD=658 m); (b) Grating-stabilized diode laser with logistic function fit (OPD=23 m). (c) The line shape as a function of observation time for the external cavity laser. In (a) and (b), the blue dots and triangles are measured $\Delta v_{\text{eff}}$ at different $T_c$, while the black line in each figure represents the curve-fit to a single logistic function.

As shown in Fig. 6(a), when $T_c$ decreases from $T_0$, $\Delta v_{\text{eff}}$ approaches $\Delta v_{\min}$, which can be considered as laser's natural linewidth. As $T_c$ increases from the "lowland", $\Delta v_{\text{eff}}$ grows rapidly due to the influence of the 1/f noise before flattening out at $\Delta v_{\max}$, which in general is dozens of times larger than $\Delta v_{\min}$. In order to cover both ends of the observation time $T_c$ for getting the complete Sigmoid curve shown in Fig. 6, two sets of data with different sampling rates were combined, one at 10 Ms/s with 1 s duration to cover the "plateau" region and the other at 1 Gs/s with 0.01 s duration to cover the "lowland" region, with each set containing 10M data points.

It is important in practice to determine the range and slope of the rapid linewidth ramping region of a laser. If we define the lower bound of the region $T_{low}$ as 10% of $\Delta v_{\max}$ and the higher bound of the region $T_{\text{high}}$ as 90% of $\Delta v_{\max}$ we obtain from Eq. (7a):

$$T_{\text{low}} = 9^{-1/p} \cdot T_0 \tag{8a}$$
$$T_{\text{high}} = 9^{1/p} \cdot T_0 \tag{8b}$$

The range of the observation time $\Delta T_c$ with rapid linewidth ramping is then:

$$\Delta T_c = T_{\text{high}} - T_{\text{low}} = (9^{1/p} - 9^{-1/p}) \cdot T_0 \tag{8c}$$

The maximum linewidth ramping slope is at $T_c = T_0$, which can be obtained by setting the derivative of Eq. (7a) to zero, as

$$R(T_0) = \frac{\Delta\nu_{\max} - \Delta\nu_{\min}}{4T_0} \cdot p \tag{8d}$$

For the effective linewidth of the external cavity laser shown in Fig. 6(a), one gets the following parameters by fitting the linewidth data at different observation times to the Sigmoid function as: $\Delta\nu_{\min} = 6.7$ kHz, $\Delta\nu_{\max} = 233$ kHz, $T_0 = 8.47$ μs, and $p = 2.26$, with $\Delta\nu_{\max}$ approximately 35 times larger than the natural linewidth $\Delta\nu_{\min}$. From Eqs. (8a-8d), one further obtains $T_{\text{low}} = 3.20$ μs, $T_{\text{high}} = 22.39$ μs, $\Delta T_c = 19.19$ μs, and $R(T_0) = 15.1$ kHz/μs. That is to say, in the region between 3.2 μs and 22.39 μs, the effective linewidth of the laser is very sensitive to the observation time, which increases at a rate of 15.1 kHz per μs increment of the observation time at $T_0 = 8.47$ μs.

The effective linewidth of the grating stabilized laser as a function of the observation time is shown in Fig. 6(b). Again, it can be perfectly fit to a logistic function, with $\Delta\nu_{\max} = 1.34$ MHz, $\Delta\nu_{\min} = 104$ kHz, $T_0 = 0.218$ μs, and $p = 1.24$, with $\Delta\nu_{\max}$ approximately 13 times larger than the natural linewidth $\Delta\nu_{\min}$. From Eqs. (8a-8d), one further obtains $T_{\text{low}} = 14.3$ ns, $T_{\text{high}} = 3.32$ μs, $\Delta T_c = 3.18$ μs, and $R(T_0) = 1.76$ MHz/μs.

Finally, the line shapes of the external cavity laser at different observation times are shown in Fig. 6(c), with which the linewidth evolution with $T_c$ can be clearly visualized.

### 3.3 Effective linewidth of a fiber laser expressed as a double Sigmoid function of observation time

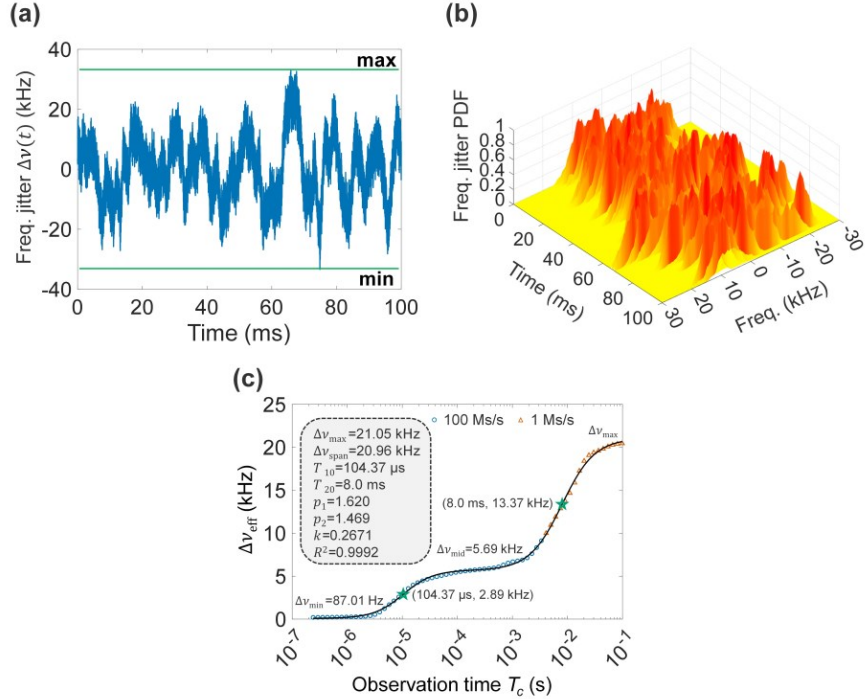

Fig. 7. Effective linewidth analysis of a NKT fiber laser. (a) Frequency fluctuation data $\Delta\nu(t)$ obtained with a sampling rate of 100 Ms/s and an OPD of 658 m; (b) 3D PDF plot of laser frequency fluctuations; (c) The effective linewidth data (blue circles and brown triangles) vs. observation time fitted to the superposition of two Sigmoid functions (black line).

To verify the generality of the Sigmoid function presentation of the linewidth of single-frequency lasers, we measured an ultra-narrow linewidth fiber laser made by NKT (Koheras BASIK E15) operating at 1550.12 nm, with an output power of 13 dBm and a specified natural linewidth of 100 Hz. Fig. 7(a) shows the measured laser frequency fluctuation data $\Delta v(t)$ vs. time obtained using the setup of Fig. 1, which were taken with an OPD of 658 m and a sampling rate of 100 Ms/s for a duration of 0.1 s. The laser frequency fluctuation range during this period is ±33 kHz. Fig. 7(b) shows the 3D PDF of the laser frequency jitter, obtained with the same procedure as in Figs. 4(b) and 4(d).

Finally, Fig. 7(c) shows the linewidth vs. observation time $T_c$ obtained using the procedure described in Fig. 3. In order to have sufficient range of the observation time $T_c$, two sets of $\Delta v(t)$ data obtained with sampling rates of 1 Ms/s for a duration of 10 s and 100 Ms/s for a duration of 0.1 s were collected and combined. In processing the data, the cutoff frequencies $f_c$ of the high-pass filter were taken from 10 MHz to 10 Hz, corresponding to the observation time from 100 ns to 100 ms.

It is interesting to note that the linewidth vs. the observation time can be perfectly fit (with the goodness of fit = 0.9992) to the superposition of two Sigmoid functions

$$\Delta v_{\text{eff}}(T_c) = \Delta v_{\max} + \frac{\Delta v_{1\min} - \Delta v_{1\max}}{1 + (T_c/T_{10})^{p_1}} + \frac{\Delta v_{2\min} - \Delta v_{2\max}}{1 + (T_c/T_{20})^{p_2}}$$
$$= \Delta v_{\max} - \Delta v_{\text{span}} \left[ \frac{k}{1+(T_c/T_{10})^{p_1}} + \frac{1-k}{1+(T_c/T_{20})^{p_2}} \right] \quad (9a)$$

where the combined maximum and minimum linewidths are

$$\Delta v_{\max} = \Delta v_{1\max} + \Delta v_{2\max} \quad (9b)$$
$$\Delta v_{\min} = \Delta v_{1\min} + \Delta v_{2\min} = \Delta v_{\max} - \Delta v_{\text{span}} \quad (9c)$$
$$\Delta v_{\text{span}} = \Delta v_{\max} - \Delta v_{\min} \quad (9d)$$
$$k = \frac{\Delta v_{1\max} - \Delta v_{1\min}}{\Delta v_{\text{span}}} = 1 - \frac{\Delta v_{2\max} - \Delta v_{2\min}}{\Delta v_{\text{span}}} \quad (9e)$$

All the parameters in Eq. (9a) can be obtained from the curve fitting, which are listed in Fig. 7(c), with the natural linewidth of the laser being 87.01 Hz, and the maximum linewidth being 21.05 kHz. As can be seen from Fig. 7(c) that in addition to the plateau with a height about 21.05 kHz, the $\Delta v_{\text{eff}}(T_c)$ curve has an intermediate plateau with a height $\Delta v_{\text{mid}}$

$$\Delta v_{\text{mid}} = \Delta v_{\max} - (1 - k) * \Delta v_{\text{span}} \quad (9f)$$

which has a value of 5.69 kHz for the NKT fiber laser. This intermediate plateau is likely due to the frequency locking loop in the laser's control system. Therefore, the double Sigmoid expression of the laser linewidth can be viewed as an indication of a laser frequency control or a RIN reduction loop in the laser system with a time constant relating to the corresponding observation time.

### 3.4 The Effective Linewidth of Desk-top Lasers expressed as Double Sigmoid Functions of Observation time

To further validate the generality of the Sigmoid function presentation of the laser linewidth vs. observation time, we measured two desk-top external cavity tunable semiconductor lasers, which are widely commercially available. The Yenista laser (TUNICS-T100SHP) operates at a fixed wavelength of 1550 nm with a driving current of 198 mA, an output power of 10 dBm and a nominal linewidth of 400 kHz specified by the manufacturer. The Newport laser (TLB-8800-HSH-CL) also operates at a fixed wavelength of 1550 nm in constant current mode with a control current of 200 mA, an output power of 11 dBm. The nominal linewidth is not specified by the manufacturer.

Fig. 8 shows the frequency fluctuation data $\Delta v(t)$ of the two lasers measured at a sampling rate of 100 Ms/s with a duration of 0.1 s, with the OPD of 658 m in the interferometer of Fig. 1 (corresponding to a frequency resolution of 7 Hz). It can be seen from Figs. 8(a) and 8(b) that the Yenista laser has a frequency fluctuation range of 23.7 MHz within 0.1 s, while the Newport laser has a range of 557.5 MHz during same time period.

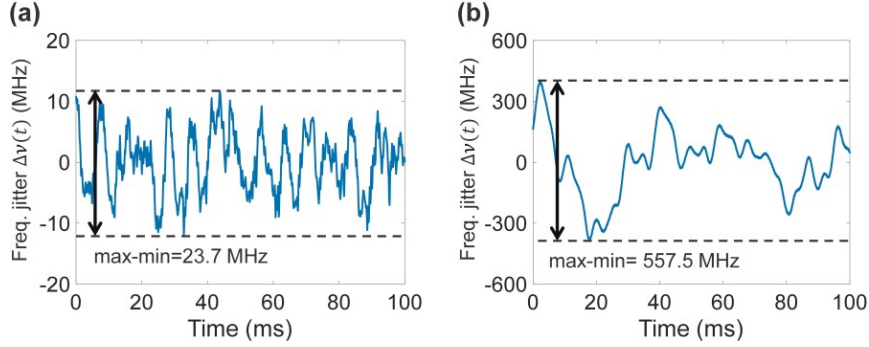

Fig. 8. Frequency fluctuations $\Delta\nu(t)$ of two desk-top tunable lasers measured with an OPD of 658 m. (a) Yenista laser (Model TUNICS-T100SHP); (b) Newport laser (Model TLB-8800-HSH-CL).

Again, by applying the TDLA procedure described in Fig. 3, we obtain the effective linewidth vs. observation time for each laser, as shown in Fig. 9(a) and Fig. 9(b). In order to have sufficient range of the observation time $T_c$, two sets of $\Delta\nu(t)$ data with sampling rates of 10 Ms/s for a duration of 1 s and 1 Gs/s for a duration of 10 ms were collected and combined for the Yenista laser in Fig. 9(a). In processing the data, the cutoff frequencies $f_c$ of the high-pass filter were taken from 100 MHz to 10 Hz, corresponding to the observation time $T_c$ from 10 ns to 100 ms. The effective linewidth $\Delta\nu_{\text{eff}}$ corresponding to each $T_c$ was then obtained by first calculating the PDF for each $T_c$ and getting its 3dB width (the blue circles and orange triangles). We find that the $\Delta\nu_{\text{eff}}$ vs. $T_c$ data can be perfectly fitted to a double Sigmoid function of Eq. (9a), with a goodness of fit of 0.9985, as anticipated because this laser also has an internal frequency control loop. All the parameters in this double Sigmoid function are listed in Fig. 9(a).

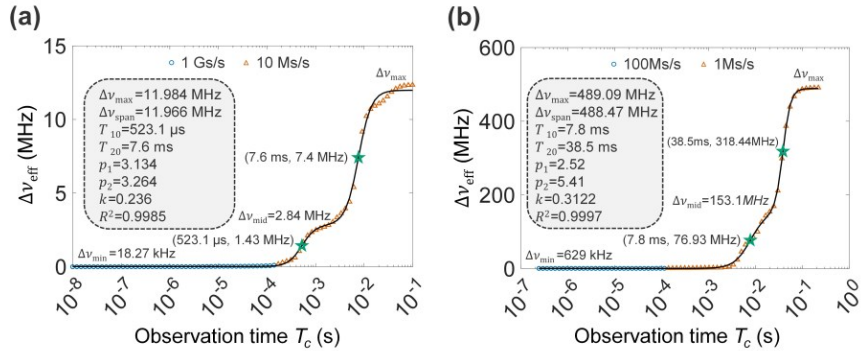

Fig. 9. The effective linewidths of the Yenista laser (a) and the Newport laser (b) as a function of observation time (blue circles and brown triangles) obtained with TDLA, with the corresponding curve-fitting to the double Sigmoid functions (black line).

It can be seen from Fig. 9(a) that the minimum and the maximum linewidths are 18.27 kHz and 11.98 MHz, respectively, located in the "lowland" and high "plateau" regions, with the maximum linewidth 656 times the minimum. Similar to Fig. 7(c), there is also an intermediate "plateau" region with a linewidth around 2.84 MHz in the vicinity of $T_c = 1$ ms, which may relate to the time constant of the laser's internal frequency control or RIN reduction loop.

The effective linewidth as a function of observation time of the Newport laser is shown in Fig. 9(b). Two sets of data with sampling rates of 1 Ms/s and 100 Ms/s were combined to get sufficient coverage of $T_c$ from 100 ns to 100ms. As expected, the effective linewidth can also be perfectly fitted to a double Sigmoid function of Eq. (9a), with a goodness of fit of 0.9997. As can be seen in Fig. 9(b), the laser has a maximum and minimum effective linewidth of 489 MHz and 629 kHz, respectively, located at the high "plateau" and "lowland" regions, respectively, with the maximum linewidth 777 times larger than the minimum. Although not as obvious as that in Fig. 9(a), there also is an intermediate "plateau" region with a linewidth around 153 MHz, located in the vicinity of 20 ms.

### 3.5 Validation our Results with Accepted Criterion and Method

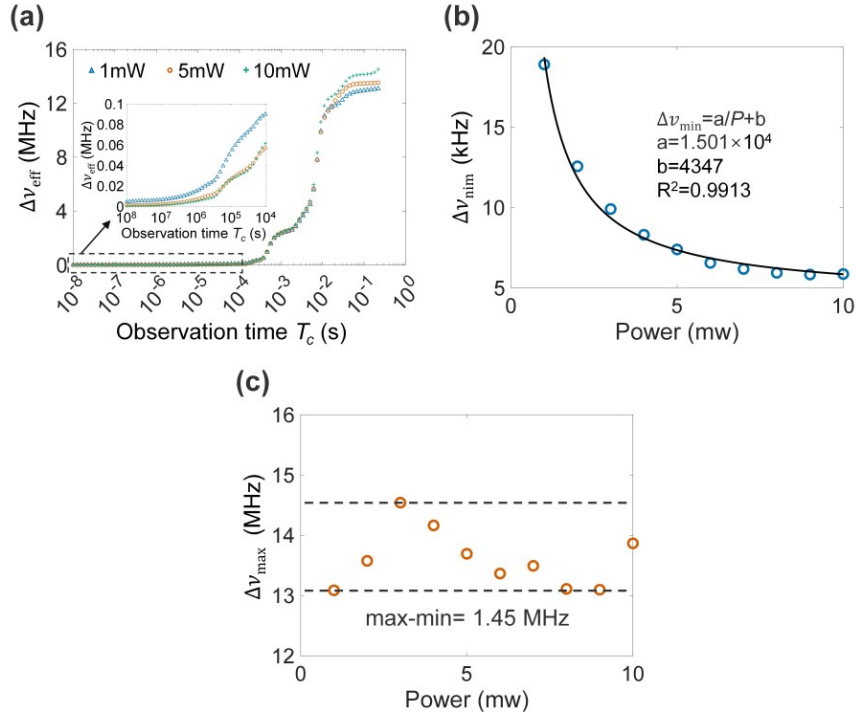

Fig. 10 (a) Effective linewidth $\Delta\nu_{\text{eff}}$ of the Yenista laser at different optical powers measured with an OPD of 658 m. (b) The natural linewidth $\Delta\nu_{\text{min}}$ of the laser obtained from the Sigmoid function fit vs. optical power. (c) The maximum linewidth $\Delta\nu_{\text{max}}$ of the laser obtained from the Sigmoid function fit vs. optical power.

As discussed previously, $\Delta\nu_{\text{min}}$ obtained from the Sigmoid function fit corresponds to the natural linewidth of a laser, which is expected to be inversely proportional to the optical power of the laser according to the STH formula [58]. Fig. 10(a) shows the measured $\Delta\nu_{\text{eff}}$ vs. $T_c$ curve of the Yenista laser corresponding to three different laser powers by adjusting the laser driving current. It can be seen from Fig. 10(a) that both $\Delta\nu_{\text{min}}$ and $\Delta\nu_{\text{max}}$ are dependent on the laser output power. Fig. 10(b) shows the $\Delta\nu_{\text{min}}$ of the Yenista laser at different optical powers determined by fitting the double Sigmoid function of the laser, which indeed has the inverse proportionality with the optical power, with an excellent goodness of fit of 0.9913. Note that the bias term b obtained from curve fitting in the figure is due to the influence of the 1/f noise [17], which is independent of the optical power [58]. On the other hand, the dependency of $\Delta\nu_{\text{max}}$ on optical power is scattered, as shown in Fig. 10(c). It is interesting to note from Fig. 10(a) that the effective linewidths at other observation times are almost independent of the optical power.

To further validate our TDLA results and the Sigmoid expression, the well accepted SDLA method is used to obtain $\Delta \nu_{eff}$ at different observation times which used the same set of raw data as that of the TDLA taken with the Yenista laser operating at a wavelength of 1550nm with an output power of 10 mw. Fig. 11(a) shows the PSD $S_{\Delta\nu}(f)$ of the laser frequency noise, in which two sets of data with different sampling rates were combined and plotted, with the frequency range from $10^0$ to $10^5$ taken with a sampling rate of 1 Ms/s, and from $10^5$ to $10^8$ taken with a sampling rate of 1 Gs/s. The black dashed line is the β-separation line $[S_{\Delta\nu}(f) = 8ln2 * f/\pi^2]$, which divides the spectrum into two areas, one contributes to the linewidth and the other defines the amplitude of white noise $h_{white}$ corresponding to the minimum or natural linewidth [25, 43]. The white noise amplitude obtained by the β-separation line is $h_{white} = 6560 Hz^2/Hz$, corresponding to the natural linewidth $\Delta\nu_{min} = \pi h_{white} = 20.6 kHz$.

The effective linewidth of the laser can be calculated with the spectral integration below [25,42,44]:

$$\Delta\nu_{eff}(T_c) = 2\sqrt{2ln2}\sqrt{\int_{1/T_c}^{\infty} S_{\Delta\nu}(f)df} \tag{10a}$$

$$\approx 2\sqrt{2ln2}\sqrt{\int_{1/T_c}^{\infty} H\left[S_{\Delta\nu}(f) - 8ln2 * \frac{f}{\pi^2}\right] S_{\Delta\nu}(f)df} \tag{10b}$$

where $H(x)$ is a step function: $H(x) = 1$ if $x \geq 0$ and $H(x) = 0$ if $x < 0$. Fig. 11(b) shows the results of the effective linewidth of the Yenista laser obtained using the SDLA method as a function of $T_c$. It is evident that the SDLA results at different observation time $T_c$ are almost identical to those of the TDLA, which can also be perfectly fitted to a logistic function.

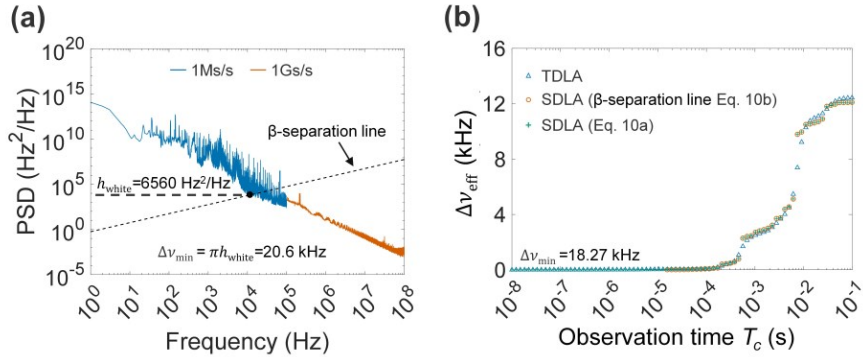

Fig. 11. (a) PSD of frequency noise of the Yenista laser. (b) The comparison of the linewidths of the Yenista laser obtained with the SDLA and TDLA methods.

Note that both the SDLA and TDLA methods can be used to obtain laser's effective linewidth for any given observation time with consistent results. However, the SDLA requires long-term measurement in order to obtain PSD of sufficiently high resolution. In addition, for each observation time, two or more sets of $\Delta\nu(t)$ data must be taken with different sampling rates, which must be combined to cover the whole required spectral range, as shown in Fig. 3(a). Finally, PSD obtained with data taken at different times needs to be averaged in order to reduce the measurement uncertainty. On the other hand, the TDLA method is more straightforward and less time consuming. For a given observation time, only one set of $\Delta\nu(t)$ data needs to be taken with a single sampling rate, which can be quickly processed to obtain PDF for obtaining the linewidth.

## 4. Summary and discussions

Table 1. Commercial single-frequency lasers characterized by a single or double Sigmoid functions

| Manufacturer | Model | Laser Type | Single Sigmoid | Double Sigmoid | |
|---|---|---|---|---|---|
| Pure Photonics | PPCL550 | External cavity tunable laser | $\Delta\nu_{max}$=233 kHz<br>$\Delta\nu_{min}$=6.71 kHz<br>$T_0$=8.47 μs<br>$p$=2.26 | ---- | |
| APEX | AP2051A | Grating stabilized diode laser in AP2051A | $\Delta\nu_{max}$=1.34 MHz<br>$\Delta\nu_{min}$=104 kHz<br>$T_0$=0.218 μs<br>$p$=1.24 | ---- | |
| HAN'S Laser | RP-MR-C34-B080-02-A | DFB laser module | $\Delta\nu_{max}$=61.1 kHz<br>$\Delta\nu_{min}$=1.08 kHz<br>$T_0$=11 μs<br>$p$=1.074 | ---- | |
| Newport | TLB-8800-HSH-CL | External cavity tunable laser | ---- | $\Delta\nu_{max}$=489.09 MHz<br>$\Delta\nu_{span}$=488.47 MHz<br>$\Delta\nu_{mid}$=153.1 MHz<br>$\Delta\nu_{min}$=629 kHz<br>$k$=0.3122 | |
| | | | | $T_{10}$=7.8 ms<br>$p_1$=2.52 | $T_{20}$=38.5 ms<br>$p_2$=5.41 |
| Yenista | TUNICS-T100SHP | External cavity tunable laser | ---- | $\Delta\nu_{max}$=11.984 MHz<br>$\Delta\nu_{span}$=11.966 MHz<br>$\Delta\nu_{mid}$=2.84 MHz<br>$\Delta\nu_{min}$=18.27 kHz<br>$k$=0.236 | |
| | | | | $T_{10}$=523.1 μs<br>$p_1$=3.134 | $T_{20}$=7.6 ms<br>$p_2$=3.264 |
| NKT | Koheras BASIK E15 | Fiber laser | ---- | $\Delta\nu_{max}$=21.06 kHz<br>$\Delta\nu_{span}$=21.04 kHz<br>$\Delta\nu_{mid}$=5.65 kHz<br>$\Delta\nu_{min}$=22.94 Hz<br>$k$=0.2675 | |
| | | | | $T_{10}$=103.97 μs<br>$p_1$=1.676 | $T_{20}$=8.0 ms<br>$p_2$=1.463 |
| Connet | CoSF-D-Er-M-1550-PM-FA | Fiber laser | ---- | $\Delta\nu_{max}$=23.07 kHz<br>$\Delta\nu_{span}$=22.66 kHz<br>$\Delta\nu_{mid}$=9.47 kHz<br>$\Delta\nu_{min}$=409 Hz<br>$k$=0.3997 | |
| | | | | $T_{10}$=127.04 μs<br>$p_1$=1.50 | $T_{20}$=9.1 ms<br>$p_2$=1.69 |

We have discovered that the linewidths of single-frequency lasers can be expressed as the sum of one or more Sigmoid functions of observation time, which have filled the long standing void for a general mathematical expression to describe the laser linewidth since the invention of the laser in 1960. We developed a sine-cosine optical frequency detection system and the time domain statistical linewidth analysis method to evaluate laser linewidths over different observation times, which have been used to confirm the generality of the Sigmoid expression with all seven (7) single-frequency lasers we measured, with laser types ranging from external cavity laser, to grating stabilized diode laser, to fiber laser, and to DFB laser, as shown in Table 1. Our OFD system has a frequency measurement resolution on the order of few hertz and can be improved to subhertz by simply increasing the fiber delay in the OFD system. We have also validated our natural linewidth measurement results with the STH formula and the associated Sigmoid expressions with the accepted SDLA methods.

The laser linewidth expression as a single Sigmoid function of observation time can be understood by considering that two statistical linewidth broadening processes are involved, one is the frequency jitter due to the spontaneous emission noise, which contributes to the natural linewidth over short observation times. The other one is the relatively slow frequency variations caused by the laser cavity fluctuations due to the temperature, vibration, carrier density and/or refractive index fluctuations, which contributes to the linewidth broadening over longer observation times with larger variations.

The laser linewidth expression as a double Sigmoid function of observation time may indicate that in addition to the two statistical linewidth broadening processes, a third process may be involved to affect the laser linewidth, such as a feedback loop inside the laser for stabilizing the laser frequency or reducing the RIN, or an additional gain medium inside the laser cavity with a different pump.

It is entirely possible for some lasers to have linewidths expressed as the sum of more than two Sigmoid functions of observation time if additional feedback loops or control mechanisms are involved, including temperature, cavity length, or pump control loops. The time constants of these control loops may be closely related with the observation times corresponding to the intermediate "plateaus".

The Sigmoid expression contains the complete information of the linewidths of a single-frequency laser over different measurement durations, which can be used to quickly assess the coherent properties of the laser for optimizing interferometric systems designs, as well as improving the laser performance, and therefore is attractive for a wide range of applications involving interferences, such as distributed acoustic sensing (DAS) and gravitational wave detections. We strongly believe that the laser and optical sensing/measurement community can be benefited significantly by adopting the elegant Sigmoid expressions reported in this paper to fully characterize the linewidths of single-frequency lasers for supporting a wide range of applications.

**Funding.** National Natural Science Foundation of China (12004092); Natural Science Foundation of Hebei Province (F2021201013, F2019201019); Science and Technology Project of Hebei Education Department (QN2020259); Supported by Interdisciplinary Research Program of Natural Science of Hebei University (DXK202204); Key R & D project of Hebei Province (20542201D); Advanced Talents Program of Hebei University (521000981006, 521000981203); Internal R&D funding of NuVision Photonics, Inc.

**Acknowledgments.** We thank Prof. Ting Feng for fruitful discussions on the delayed heterodyne linewidth measurement method and Dr. James Chen of Inline Photonics for the help in building the OFD system.

**Data availability.** Data is available from corresponding author upon reasonable requirements.

**Supplemental document.** See Supplement 1 for supporting content.

**Author contributions**: X.S.Y. initiated and guided the project, designed the optical frequency detection system and supervised it' s assembly and test, proposed the idea of using a Sigmoid function to represent the "S" shaped linewidths of lasers observed in the experiment, and revised the manuscript draft. X.M. conducted all measurements and data processing, and drafted the initial manuscript. X.S.Y. and X.M. discussed the results and edited the manuscript together.

**Conflict of interest**: The authors declare no competing interests.

# SINGLE-FREQUENCY LASERS' LINEWIDTH ELEGANTLY CHARACTERIZED WITH SIGMOID FUNCTIONS OF OBSERVATION TIME: SUPPLEMENTAL DOCUMENT

**Limitations and uncertainties of the OFD system with the TDLA approach**

**I. Frequency resolution limited by the resolution of DAQ card**

From [1], the frequency measurement resolution of the OFD system due to the resolution of the DAQ card is

$$\delta\nu_{\text{ADC}} = FSR/(2 \cdot 2^M) = 1/(2^{M+1}\tau) \tag{S1}$$

where FSR represents the free spectral range of the interferometer, $M$ represents the number of bits of the analog-to-digital converter (ADC) of the DAQ card. In order to accurately measure the laser linewidth, $\delta\nu_{\text{ADC}}$ must be much finer than the laser linewidth to be measured. For example, if the acquisition card has a resolution of 16 bits, the frequency resolution corresponding to the OPD of 658 m is 7 Hz (assuming an effective resolution of 15 bits), which is much finer than the natural linewidths of the LUTs in our measurements on the order of tens Hz to MHz. Proportionally longer OPD can be used for LUTs with narrower natural linewidths.

**II. The maximum detectable frequency variation rate**

In addition, the measurement system must be sufficiently fast to capture the fastest laser frequency jitters. As discussed in [1], the speed of the corresponding signal is proportional to the OPD $\tau$. Considering that our DAQ has a maximum sampling rate $R_s$ of 1 Gs/s, the maximum frequency variation rate $\gamma_{\text{max}}$ can be measured is $\gamma_{\text{max}} = R_s/(2\tau)$ [1]. Corresponding to the OPD of 658 m ($\tau = 2.2\ \mu$s), the resulting $\gamma_{\text{max}}$ is 228 THz/s. In order to increase $\gamma_{\text{max}}$, one may reduce the OPD $\tau$ or increase the sampling rate $R_s$. It is advisable to first determine the maximum frequency changing rate of the LUT and then determine the best suitable OPD and $R_s$.

**III. Measurement uncertainty due to the system noise**

As discussed in section 2.2, the system noise can contribute to the measurement uncertainty and must be subtracted from the effective linewidth measurement data following the procedure described in Fig. 3. Unfortunately, residual system noise contribution may still present and therefore it is important to find out the linewidth measurement uncertainty even when the subtraction procedure of the system noise equivalent linewidth $\Delta\nu_{n,V}$ is applied.

The equivalent optical frequency fluctuation $\Delta\nu_n(t)$ due to the system noise can be evaluated with Eqs. (5b) and (5c), which must be obtained prior to obtaining $\Delta\nu_{n,V}$. Fig. S1 shows an example of the three photo-voltages of the interferometric signal when the NKT laser (Koheras BASIK E15) entering the OFD system of Fig. 1, with the blue curves including the contributions of both the laser frequency fluctuation and the system noise, which are used to compute $\Delta\theta(t)$ using Eq. (3), after parameters $C_j$, $B_j$, and $\beta_j$ have been obtained using the calibration procedure described in section 2.1. The red curves represent the three photo-voltages $\widetilde{V}_j(t)$ obtained using $\widetilde{\Delta\theta}(t)$, the low-pass filtered $\Delta\theta(t)$, in Eq. (5b) to remove the contributions of rapid laser frequency jitter and system noise. Finally, the green curves are the three photo-voltages $V'_{nj}(t)$ calculated using Eq. (5c), representing the system noise $\delta V_{nj}(t)$ superimposed onto the interference signals, which are further used in Eq. (3) to compute the corresponding $\Delta\nu_n(t)$ due to the system noise alone. The cutoff frequency $f_{LPF}$ of the low-pass filter in Fig. S1 is chosen to be 5 Hz while the data is taken with a sampling rate $R_s$ of 1 Ms/s. Different cutoff frequencies $f_{LPF}$ should be chosen for data taken with different sampling rates $R_s$ for optimal results.

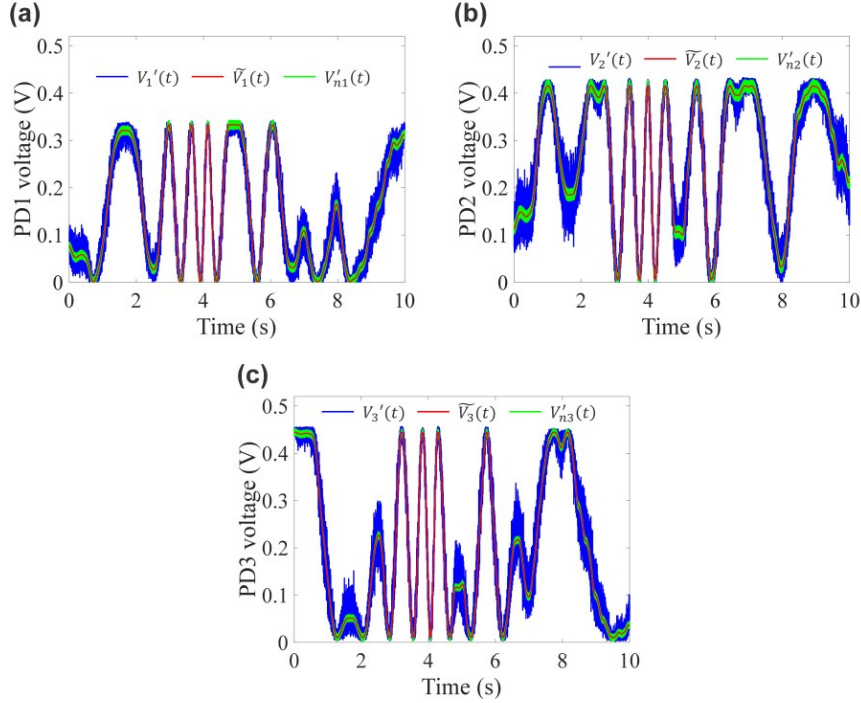

Fig. S1. The three photo-voltages (a), (b), and (c) of the interference signal. The blue curve is the original interference signal $V_j'(t)$, including the laser frequency fluctuations and the system noise, the red curve is the jitter-free interference signal $\widetilde{V}_j(t)$, where a low-pass 3rd-order Butterworth filter with a cutoff frequency of 5 Hz is used to remove the frequency jitter and system noise. The green curve shows the system noise $V'_{nj}(t)$ alone superimposed on to the filtered interference signal.

Fig. S2 shows the $\Delta v_{n,V}$ of the measurement system calculated with $\delta V_{n1}(t)$, $\delta V_{n2}(t)$ and $\delta V_{n3}(t)$ at four different sampling rates of 1 Gs/s, 100 Ms/s, 10 Ms/s, and 1 Ms/s when the LUT is turned off while the OPD is set at 658 m. In the calculation, the following typical parameters in Eqs. 3, (5b) and (5c) are assumed to compute $\Delta v_n(t)$: $C_j = 0.25$, 0.3, 0.31 V, $B_j = 0.21$, 0.26, 0.28 V, $\beta_j = 0$, $-118.9°$, $122.4°$, $j = 1,2,3$, and $\widetilde{\Delta\theta}(t) = 0$, which are obtained using the calibration process described in section 2.1 with a 1550 nm laser outputting a power of 7.5 dBm into the OFD system. $\Delta v_{n,V}$ can be obtained by following the procedure in Fig. 3 after $\Delta v_n(t)$ is obtained. It can be seen that the $\Delta v_{n,V}$ corresponding to the 1 Gs/s sampling rate is 825 Hz, notably higher than those of other sampling rates. According to [2, 3], the relative uncertainty of $\Delta v_{n,V}$ from curve fitting is between $10^{-2}$ to $10^{-5}$.

If the laser effective linewidth $\Delta v_{eff}$ at a particular observation time is much larger than $\Delta v_{n,V}$, the effect of this $\Delta v_{n,V}$ on the laser $\Delta v_{eff}$ is negligible, as the case of Figs. (5a) and (5b). However, if $\Delta v_{n,V}$ is comparable with $\Delta v_{eff}$, the $\Delta v_{n,V}$ subtraction must be applied. Finally, if $\Delta v_{n,V}$ is larger than $\Delta v_{eff}$, the $\Delta v_{n,V}$ subtraction may not be sufficient and large measurement uncertainty may result.

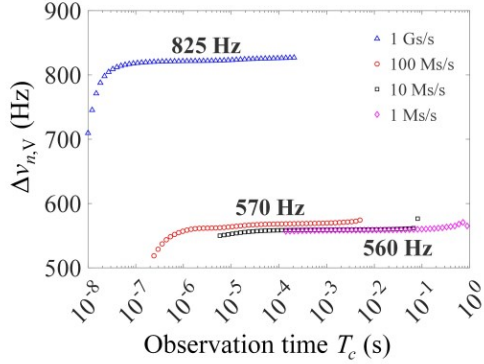

Fig. S2. System noise equivalent linewidths $\Delta v_{n,V}$ as a function of observation time at different sampling rates measured with an OPD of 658 m.

Fig. S3(a) shows the results of 20 repetitive $\Delta v_{\text{eff}}$ measurement of the Yenista laser operating at a fixed wavelength of 1550 nm as a function of observation time $T_c$, with error bars representing the standard deviation. Fig. S3(b) shows the corresponding relative uncertainty of $\Delta v_{\text{eff}}$ at different $T_c$, with the worst relative uncertainty of 2.62%. This small measurement uncertainty is due to the fact that the $\Delta v_{n,V}$ of the measurement system is much smaller than the laser $\Delta v_{\text{eff}}$ over all observation times $T_c$. In the measurement, the output power of the Yenista laser was set to 10 dBm and the sampling rate of the DAQ was set at 1 Ms/s and 1 Gs/s, as indicated inside Fig. S3(a).

Fig. S3(c) shows the results of 20 repetitive $\Delta v_{\text{eff}}$ measurements of the NKT fiber laser over different observation times $T_c$ with the $\Delta v_{n,V}$ subtraction applied, with error bars representing the standard deviation, while Fig. S3(d) shows the relative linewidth measurement uncertainty derived from Fig. S3(c). It can be seen that a relative linewidth uncertainty for the natural linewidth $\Delta v_{\min}$ is as large as 21%, despite of $\Delta v_{n,V}$ subtraction because 1) $\Delta v_{n,V}$ is larger than $\Delta v_{\min}$; and 2) the small denominator when computing the relative uncertainty due to the small $\Delta v_{\min}$. Nevertheless, the near perfect Sigmoid curve-fit of the linewidth as a function of observation time is not compromised due to the averaging effect of the curve fitting. In the measurement, the output power of the fiber laser was set to 13 dBm, the center wavelength was set at 1550.12 nm, and the sampling rate of the DAQ was set at 100 Ms/s and 1 Ms/s, as indicated inside Fig. S3(c).

It is therefore important to find out the impact of the $\Delta v_{n,V}$ on the measurement uncertainty of $\Delta v_{\text{eff}}$. Fig. S4 shows the numerically simulated relative linewidth measurement error (RLWE) at different laser to noise linewidth ratios, more specifically the ratio of laser's true linewidth $\Delta v_l$ over the system noise equivalent linewidth $\Delta v_{n,V}$ ($\Delta v_l/\Delta v_{n,V}$). In the simulation process, the system noise in Fig. 1 at a particular sampling rate was first taken, with its PDF calculated as $P_n(v)$ and the corresponding system noise equivalent linewidth as $\Delta v_{n,V}$. A LUT is assumed to have a frequency noise PDF of $P_l(v)$, which is Gaussian white noise with a linewidth of $\Delta v_l$. After adding the system noise to the LUT's frequency noise, the combined linewidth can be obtained as $\Delta v_{l+n,V}$. After subtracting the contribution of $\Delta v_{n,V}$ following the procedure of Fig. 3, the final Voigt linewidth of the laser can be obtained as $\Delta v_{\text{eff}}$. The RLWE is calculated as

$$RLWE = (\Delta v_{\text{eff}} - \Delta v_l)/\Delta v_l \tag{S2}$$

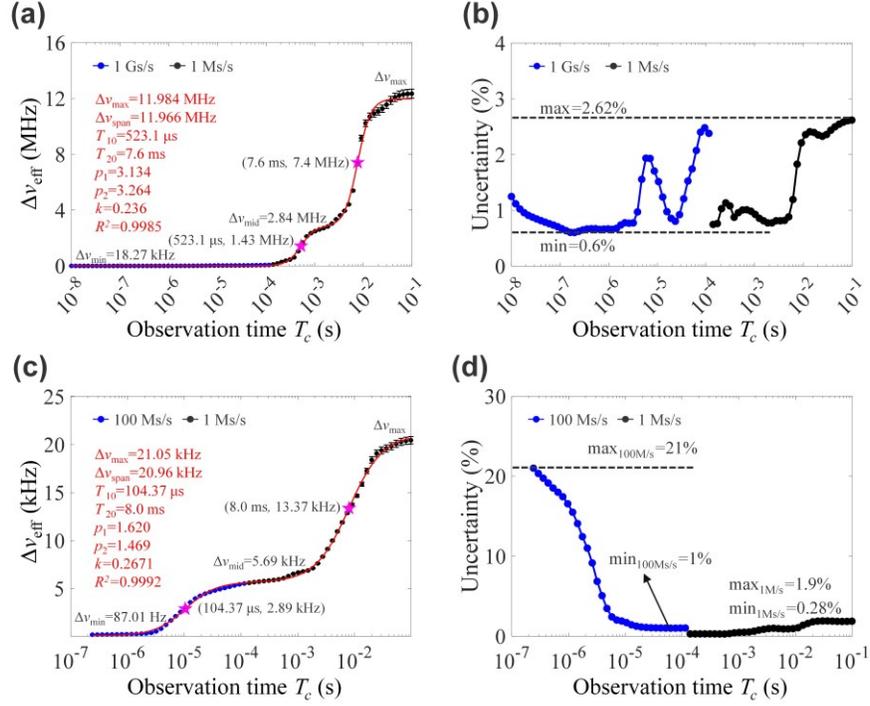

Fig. S3. 20-time repeatability measurement of the effective linewidth using the OFD of Fig. 1 (OPD=658 m) and the TDLA method. (a) Yenista laser at different observation times obtained by combining two sets of data with sampling rates of 1 Gs/s and 1 Ms/s, with the error bars representing the standard deviation; (b) The relative linewidth uncertainty of the Yenista laser derived from the data in (a); (c) Effective linewidth and error bars (standard deviation) of the NKT fiber laser at different observation times obtained by combining two sets of data with sampling rates of 100 Ms/s and 1 Ms/s; (d) The relative linewidth uncertainty derived from (c).

As shown in Figs. S4(a) and S4(b), when $\Delta v_l/\Delta v_{n,V}$ is greater than 10, the RLWE is stable within ±0.2%. When $\Delta v_l/\Delta v_{n,V}$ is less than 10, the RLWE increases rapidly and gradually approaches to 10%, which is consistent with the uncertainty result of Fig. S3(d).

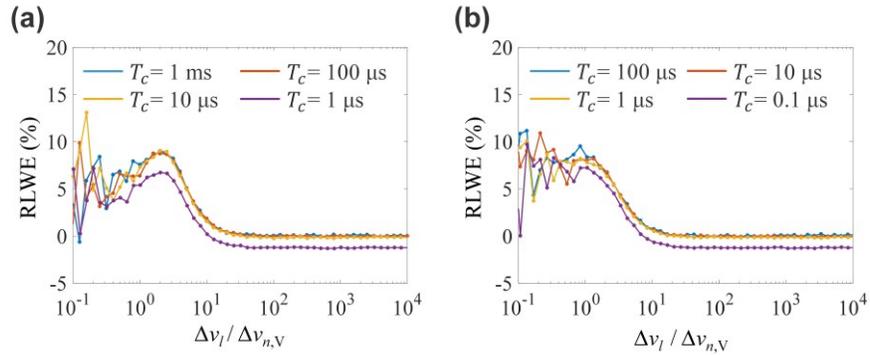

Fig. S4. Simulated relative linewidth measurement error (RLWE) as a function of $\Delta v_l/\Delta v_{n,V}$ over different observation times (OPD=658 m). (a) At a sampling rate of 100 Ms/s with $\Delta v_{n,V} = 570$ Hz. (b) At a sampling rate of 1 Gs/s with $\Delta v_{n,V} = 825$ Hz.

**IV. The range of observation time limited by the sampling rate and time**

For statistical analysis, it is necessary for the signal to contain sufficient number of valid sampling points to obtain the PDF of frequency fluctuations for a given observation duration $T_c$. The sampling rate $R_s$, sampling time $\Delta T_s$, and the total number of sampling points $M_s$ are related by:

$$M_s = R_s \Delta T_s \tag{S3}$$

In this paper, $M_s$ during data acquisition is fixed at 10 million points. Therefore, both the sampling rate and time must be chosen accordingly. For wide linewidth lasers, larger $R_s$ is required to capture fast frequency variations, and $\Delta T_s$ must be reduced proportionally. While for narrow linewidth lasers, smaller $R_s$ is required, and $\Delta T_s$ should be increased accordingly. We found during data processing that for obtaining linewidths with acceptable uncertainty, the observation time should be in the range of:

$$10/R_s \leq T_c \leq \Delta T_s/100 \tag{S4}$$

For lasers with a large linewidth span $\Delta\nu_{\max} - \Delta\nu_{\min}$ over the observation time, two sampling rates $R_{s1}$ and $R_{s2}$ and the corresponding sampling times $\Delta T_{s1}$ and $\Delta T_{s2}$ are generally chosen to cover both ends, with $R_{s1} > R_{s2}$ and $\Delta T_{s1} < \Delta T_{s2}$. The corresponding range of observation time is

$$10/R_{s1} \leq T_c \leq \Delta T_{s2}/100 \tag{S5}$$

For example, for a sampling rate and time of 1 Gs/s and 0.01 s, the range of $T_c$ is $10^{-8} < T_c < 10^{-4}$ s. For a measurement requiring two sampling rates 100 Ms/s and 1 Ms/s, with sampling times 0.1 s and 10 s, respectively, the range of $T_c$ is $10^{-7} < T_c < 10^{-1}$ s. All the linewidth results as a function of observation time in Figs. 6, 7, 9, and 10 in this paper are largely compliant with this rule.

## V. Measurement uncertainty due to system temperature variations

Finally, the temperature variations may affect the OPD of the Michelson interferometer in the OFD system in Fig. 1, causing frequency measurement uncertainties. Therefore, it is important to evaluate its impact on the laser linewidth measurement. Fig. S5 shows the measured temperature inside the thermally insulated enclosure containing the interferometer, which varied 1.02 °C over a period of 60 minutes, corresponding to a temperature changing rate approximately 2.8×10$^{-4}$ °C/s. From Eq. (4), the frequency detection error $\delta\nu$ due to the OPD delay variation $\delta\tau$ can be obtained as:

$$\delta\nu/\nu = -\delta\tau/\tau = -C_n \cdot \Delta\text{Temp} \tag{S6}$$

where $C_n$ is the temperature coefficient of the optical fiber with $C_n \approx 0.82 \times 10^{-5}/°C$ [4], and $\Delta\text{Temp}$ is the temperature variation. For the observation times of 1 $\mu$s, 1 ms, and 0.1 s, the corresponding $\Delta\text{Temp}$ are 2.8×10$^{-10}$, 2.8×10$^{-7}$, and 2.8×10$^{-5}$ °C, respectively, resulting in $\delta\nu$ of 0.45 Hz, 450 Hz, and 45 kHz, which are orders of magnitudes smaller than the laser frequency fluctuations at the corresponding observation times in most cases, except for the NKT fiber laser at $T_c = 0.1$ s with comparable magnitudes (see Fig. 7). Therefore, for ultra narrow linewidth lasers, active temperature control of the OPD in the OFD system is required.

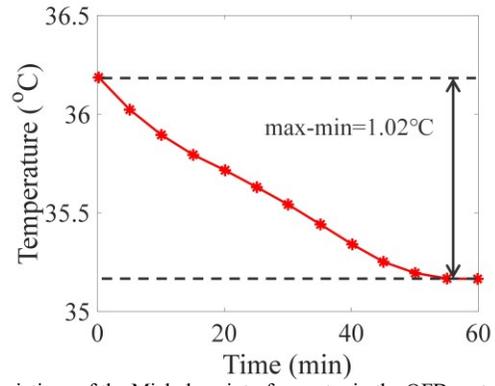
Fig. S5. Temperature variations of the Michelson interferometer in the OFD system of Fig. 1.